\documentclass[prb]{revtex4-2}
\usepackage{graphicx}
\usepackage{dcolumn}
\usepackage{bm}
\usepackage [latin1]{inputenc}
\usepackage{epstopdf}
\usepackage{color}
\usepackage{ulem}
\usepackage{multirow}

\begin{document}
\title{Stationary waves in a superfluid gas of electron-hole pairs in bilayers}

\author{D.\,V.\,Fil$^{1,2}$, S.\,I.\,Shevchenko$^3$}


\affiliation{$^1$Institute for Single Crystals, National Academy of Sciences of Ukraine,
60 Nauky Avenue, Kharkov 61072, Ukraine\\
$^2$V.N. Karazin Kharkiv National University, 4 Svobody Square, Kharkov 61022,
Ukraine\\$^3$B.~Verkin Institute for Low Temperature Physics and Engineering,\\ National
Academy of Sciences of Ukraine,\\ 47 Nauky Avenue,  Kharkov 61103, Ukraine}

\begin{abstract}

Stationary waves  in the condensate of electron-hole pairs in the $n-p$ bilayer system
are studied.  The system demonstrates the transition from a uniform (superfluid) to a
nonuniform (supersolid) state. The precursor of this transition is the appearance of the
roton-type minimum  in the collective mode spectrum. Stationary waves occur in the flow
of the condensate past an obstacle. It is shown that the roton-type minimum manifests
itself in a rather complicated stationary wave pattern with several families of crests
which cross one another. It is found that  the stationary wave pattern is essentially
modified under variation in the density of the condensate and under variation in the flow
velocity. It is shown that the pattern is formed in the main part by shortwave modes in
the case of a point obstacle. The contribution of longwave modes is clearly visible in
the case of a weak extended obstacle, where the stationary wave pattern resembles the
ship wave pattern.
   \end{abstract}

\maketitle

\section{INTRODUCTION}

Stationary waves emerge in systems in which the collective mode spectrum is different
from the linear one. The well-known example is  ship waves on a
 water surface
(see, for instance, \cite{1}).  The theory of ship waves was developed by Kelvin.
 The interest in
stationary waves was renewed recently in connection with the study of the superfluidity
of atomic Bose-Einstein condensates (BECs). Stationary waves in $^{87}$Rb BEC were
observed in \cite{e2} (see also \cite{e3}).

 The stationary wave pattern reported in \cite{e2} is
in agreement with the theoretical prediction \cite{3, 3-1, 3-2}. It was shown in \cite{3,
3-1, 3-2} that stationary waves in the atomic BEC are excited outside the Mach cone. It
differs from the case of ship waves on a surface of deep water where waves  fill the
wedge-shaped region behind the obstacle with the semiangle $\approx 19.5^\circ$
\cite{urs}. Stationary waves in a two-component atomic BEC were studied in
\cite{two1,two2}. In two-component condensates the stationary wave pattern is more
complicated due to the presence of two Mach cones. In this case the waves are not excited
in the intersection of two Mach cones.

The stationary wave pattern similar to \cite{3, 3-1, 3-2} emerges in a superfluid gas of
exciton polaritons in microcavities \cite{2,bc}. More complicated stationary wave
patterns with waves inside the Mach cone  occur in the flow past a rigid extended
obstacle \cite{p1,p2}. Exciton polaritons have very small effective mass (of the order of
$10^{-5}$ times the free-electron mass) and due to this the temperature of the transition
to the superfluid state can be quite high. The observation of stationary waves in the
flow of the condensate past an obstacle with overcritical velocities and the absence of
such waves at smaller velocities  is considered \cite{2a} as a proof for the exciton
polariton superfluidity.
 Due to the smallness of the
binding energy of Wannier-Mott excitons in the AlGaAs microcavity  in \cite{2},
superfluid behavior was observed at $T=5$ K. Later a behavior similar to \cite{2} were
observed for Frenkel exciton-polaritons in organic microcavities at room temperature
\cite{2-1}.

Exciton polaritons have a finite lifetime and laser pumping is required to support the
superfluid state.  Spatially-indirect excitons in electron-hole bilayers may have an
infinite lifetime.  In the high-density limit,  these excitons are similar to Cooper
pairs in superconductors, and the description of such systems  can be given in the
framework of a modified version \cite{4,5,6,7,8}  of the Bardin-Cooper-Schrieffer (BCS)
theory \cite{bcs}. In the low-density limit the approach based on the Keldysh wave
function \cite{kel} can be used \cite{litl,10}.  The superfluidity  of spatially indirect
excitons can be considered  as a special kind of superconductivity \cite{4,5}.
Nondissipative transport can be realized in the counterflow setup, where electric
currents in the adjacent layers are equal in modules and flow in opposite directions. The
modern state of the problem of the electron-hole condensate in bilayers as of 2018 is
reviewed  in \cite{22}.

Counterflow superconductivity in  quantum Hall bilayer systems, i. e., $n$-$n$ and
$p$-$p$ ones, with  the total filling factor of Landau levels $\nu_T=1$  was  predicted
in \cite{11,12,13}.  At $\nu_T=\nu_1+\nu_2=1$ ($\nu_i$ is the filling factor of the
$i$-th layer), the concentration
 of filled states (electrons) in one layer is equal to the concentration of empty states
(holes) in the zero Landau level in the other layer.

Anomalies in transport properties  caused by the electron-hole pairing were observed
  in quantum Hall bilayers \cite{17,18,19,20} (see,
also, the review \cite{21}), as well as in $n$-$p$ bilayers in a zero magnetic field
\cite{14,15}.

The possibility  of electron-hole pairing in graphene systems was initially considered
with reference to double monolayer  graphene (DMG)  \cite{g1,g2}, and then, with
reference to double bilayer graphene (DBG) \cite{bl-1}. Due to the linear dispersion of
the energy band in the monolayer graphene, electron-hole coupling in DMG is weak at all
densities of carriers. Therefore, the screening almost suppresses the electron-hole
pairing \cite{g3}. In contrast, at low density of carriers  the electron-hole coupling in
DBG is strong and the binding is not suppressed by screening. In a recent paper
\cite{bl-2}, the results of a thorough study of electron-hole pairing in DBG were
presented. The authors of \cite{bl-2} conclude that DBG can be considered as an optimum
platform for realizing and exploiting electron-hole superfluidity.   In experiment
\cite{16}, the low-temperature enhancement of the tunneling conductance in the DBG system
at matched concentrations of electrons and holes was registered. The effect is considered
as a clear signature of electron-hole pairing. The mechanism of the conductance
enhancement \cite{16} connected with the fluctuational internal Josephson effect was
proposed in \cite{ef}. The theory \cite{ef} predicts that cleaner samples with longer
disorder scattering times would demonstrate condensation of electron-hole pairs at
temperatures $T_c$ up to 50 K, compared to the record $T_c\approx 1.5$ K achieved in the
experiment \cite{16}.

AlGaAs double-well heterostructures are less promising because of the small effective
electrons mass and large dielectric constant of the matrix. The screening of the Coulomb
interaction suppresses already the electron-hole pairing  at small density of carriers.
The estimate for the maximum temperature of the superfluid transition obtained in
\cite{gaas} is of the order of 0.1 K.

Transition metal dichalcogenide (TMD) double layers are considered to be a very good
perspective for a realization of high-temperature exciton superfluidity \cite{m1,m2-1,m2,
deb, pee, m3,mr}. The parabolic bands in TMDs allow one to achieve the regime of strongly
bound electrons and holes. It is the low-density regime in a sense that the size of the
pair is much smaller than the average distance between the pairs. In contrast to GaAs
heterostructures with  double quantum wells, the effective masses of carriers in TMDs are
almost equal to each other and relatively large. Therefore, the low-density regime
corresponds to physically much larger densities. The valence band screening is negligible
in TMDs because of the large band gap, in contrast to the DBG systems. Recent observation
of a large enhancement of electroluminescence in the MoSe$_2$ - WSe$_2$ double layer
\cite{m4} confirms the condensation of electron-hole pairs in these systems at the
temperature about 100 Kelvin.

Stationary waves in a condensate of spatially-indirect excitons can be generated by the
counterflow current if it exceeds the critical current  $j_c^{\mathrm{L}}$ given by the
Landau criterion of superfluidity. Stationary waves in a superfluid gas of  bound
electron-hole pairs in quantum Hall bilayers were considered by one of the authors in
\cite{pik1}. One could expect similar behavior of stationary waves in quantum Hall
bilayers and in electron-hole bilayers in a zero magnetic field. But there is an
essential restriction for a realization of stationary waves in quantum Hall bilayers
\cite{pik1}. The point is that in quantum Hall bilayers, the uniform state becomes
dynamically unstable if the counterflow current $j$ exceeds a certain critical value
$j_c^{\mathrm{dyn}}$ \cite{ab, kr1,kr2}. In balanced quantum Hall bilayers
($\nu_1=\nu_2=1/2$) the critical current $j_c^{\mathrm{dyn}}$ coincides with the critical
current  $j_c^{\mathrm{L}}$.
 Therefore, the range of currents at which stationary waves can be excited shrinks to zero.
   If $\nu_1\ne \nu_2$, the  critical current $j_c^{\mathrm{L}}$ is smaller
   than $j_c^{\mathrm{dyn}}$ and stationary waves are excited in the range of
   currents, $j_c^{\mathrm{L}}<j<j_c^{\mathrm{dyn}}$.
   In weakly imbalances quantum Hall bilayers, the corresponding range of $j$ is quite narrow.

In this paper, we investigate stationary waves in the superfluid gas of electron-hole
pairs in $n-p$ bilayers in the absence of a magnetic field. We consider the low-density
limit  and use the coherent wave-function approach. The approach was proposed by Keldysh
\cite{kel} for a description of the coherent state of excitons in three-dimensional
systems (without spatial separation of carriers). In \cite{10,my2}, the approach
\cite{kel} was used for the analysis of the conditions of stability of the BEC of
electron-hole pairs in bilayers and the transition of such a BEC to the supersolid state.
The spectrum obtained in \cite{10} (much like the collective excitation spectrum in
quantum Hall bilayers \cite{11}) has a roton-type minimum close to the
superfluid-supersolid transition line \cite{my2, exss1,exss2}. This transition is similar
to the superfluid-supersolid transition in dipole Bose gases \cite{loz,petrov,
ss1,ss2,ss3}. It is important to clarify how the roton-type minimum reveals itself in the
stationary wave pattern.  One can expect that the pattern will be quite complicated.
 In the
general case,
 several families of stationary waves which cross one
another may emerge and the stationary wave pattern may be sensitive to a change in the
flow velocity.

In Sec. \ref{s2}, we review the approach \cite{kel} and derive the collective mode
spectrum in the moving condensate. In Sec. \ref{s3}, we present the kinetic  description
of  stationary waves and find how the wave crest pattern is modified under variation in
the density of the condensate and under variation in the flow velocity. In Sec. \ref{s4},
the dynamical approach to the description of stationary waves in the exciton BEC is
developed. We consider the $\delta$-function potential of the obstacle (the point
obstacle) and  the Gaussian potential  of the obstacle  (a weak extended obstacle). It is
shown that in the former case, the stationary wave pattern is formed in the main part by
short waves while in the latter case, the long-wave contribution becomes  visible.

\section{Coherent wave function for the exciton condensate in bilayers} \label{s2}

In this section, we review the coherent wave-function approach to the description of  BEC
of spatially indirect excitons \cite{kel,10} and derive formulas used for the calculation
of the stationary wave patterns in Secs. \ref{s3} and \ref{s4}. In \cite{10}, we found
the collective mode spectrum for the exciton BEC at rest. Here we obtain the collective
mode spectrum in the flowing condensate. The approach \cite{10} is based on the
many-particle wave function which describes the coherent state of the gas of bound
electron-hole pairs in the bilayer. The function has the form \cite{kel}
\begin{equation}\label{1}
 |\Phi\rangle=\hat{D}_\Phi|0\rangle,
\end{equation}
where
\begin{equation}\label{2}
  \hat{D}_\Phi=\exp\left(\int d {\bf
    r}_1 d{\bf
    r}_2 \Phi({\bf
    r}_1, {\bf
    r}_2,t)\psi_e^+({\bf
    r}_1) \psi_h^+({\bf
    r}_2)-H.c.\right) ,
\end{equation}
$\psi_e^+$ and $\psi_h^+$ are the electron and hole creation operators, respectively, and
the wave function $|0\rangle$ corresponds to the vacuum state (the state without
electrons and holes). The function $\Phi({\bf
    r}_1, {\bf r}_2,t)$ can be interpreted as the BEC order parameter for a gas
of bound electron-hole pairs. The radius vectors $\mathbf{r}_1$ and $\mathbf{r}_2$ are
two-dimensional ones and relate to the $n$ and $p$ layers respectively. The wave function
(\ref{1}) is a generalization of the many-particle wave function of the BEC of
structureless particles $|\Phi_{BEC}\rangle=\exp\left[\int
d\mathbf{r}(\Psi(\mathbf{r})b^+(\mathbf{r})- H.c.)\right]|0\rangle$, where
$b^+(\mathbf{r})$ is the bosonic creation operator and $\Psi(\mathbf{r})$ is the BEC
order parameter (see, for instance, \cite{10}).

Substitution of the wave function (\ref{1}) into the Schroedinger equation yields the
equation \cite{kel}
\begin{equation}\label{3}
\left(i\hbar \hat{D}_\Phi^+ \frac{\partial \hat{D}_\Phi}{\partial t}- \hat{D}_\Phi^+ H
\hat{D}_\Phi\right)|0\rangle=0,
\end{equation}
where $H$ is the Hamiltonian of the system. Strictly speaking,  Eq. (\ref{3}) can be
satisfied only approximately. The function (\ref{1}) corresponds to the description of
the     electron-hole pairing in the self-consistent approximation and  does not take
into account multi-particle correlation effects. The equation
    for the function $\Phi({\bf
    r}_1, {\bf r}_2,t)$ in the self-consistent approximation can be obtained from Eq. (\ref{3})
    (see \cite{kel}).
    This function can also be found from the condition that the variation of the expectation
    value of the grand potential,      $G=\langle \Phi |H-\mu \hat{N}|\Phi\rangle$
     ($\hat{N}$ is the pair number operator),
     with respect to $\Phi({\bf     r}_1, {\bf r}_2,t)$ is equal to zero \cite{rukin10-2}.
    The situation is similar to one for the BCS wave function \cite{bcs}. The
     $u-v$ coefficients that enter into the BCS wave function can be found
    by diagonalization of the mean-field Hamiltonian or by minimization of the
    mean-field energy.

We specify the case where the $n$ and $p$ layers are embedded into a homogeneous
dielectric matrix (a three-layer matrix is considered in Appendix \ref{aa}). The
Hamiltonian of the system has the form
\begin{eqnarray}\label{4}
    H=-\sum_{\alpha=e,h}\int d {\bf r}\frac{\hbar^2}{2
    m_\alpha}\psi^{+}_{\alpha}({\bf r})\nabla^2
    \psi_{\alpha}({\bf r})\cr
    +\frac{1}{2}\sum_{\alpha,\beta=e,h}
    \int d {\bf
    r}d{\bf
    r'}\psi^{+}_{\alpha}({\bf r})
    \psi^+_{\beta}({\bf r'})
    V_{\alpha\beta}(|{\bf r}-{\bf r'}|)
    \psi_{\beta}({\bf r'})\psi_{\alpha}({\bf
    r}),
\end{eqnarray}
where  $V_{ee}(r)=V_{hh}(r)=e^2/\varepsilon r$ and $V_{eh}(r)=-e^2/\varepsilon
\sqrt{r^2+d^2}$ are the Coulomb energies,   $m_e$ and $m_h$ are the effective masses of
electrons and  holes, and $\varepsilon$ is the dielectric constant  of the matrix.

The function $\Phi({\bf r}_1, {\bf r}_2,t)$ is sought in the form
\begin{equation}\label{5}
\Phi({\bf r}_1, {\bf r}_2,t)=\Psi(\mathbf{R}_{12},t)\phi_0(\mathbf{r}_{12}),
\end{equation}
where $\mathbf{R}_{12}=(m_e\mathbf{r}_1+m_h\mathbf{r}_2)/(m_e+m_h)$ is the center-of-mass
radius vector of  the pair, $\mathbf{r}_{ik}=\mathbf{r}_k-\mathbf{r}_i$, and
$\phi_0(\mathbf{r})$ is the ground-state wave function of the pair. The function
$\phi_0(\mathbf{r})$ satisfies  the Schroedinger equation for the isolated pair,
\begin{equation}\label{6}
\left[-\frac{\hbar^2}{2 m_*}\nabla^2
+V_{eh}(r)\right]\phi_0(\mathbf{r})=E_0\phi_0(\mathbf{r}),
\end{equation}
where $m_*=m_e m_h/(m_e+m_h)$ is the reduced mass, and $E_0$ is the ground state energy.
The function $\phi_0(\mathbf{r})$ in (\ref{5}) is normalized by the condition $\int d
\mathbf{r} |\phi_0(\mathbf{r})|^2=1$. The function $\Psi(\mathbf{R},t)$ can be considered
as the one-particle condensate wave function. This function satisfies the equation
\cite{10,40}
\begin{eqnarray}\label{7}
i \hbar \frac{\partial}{\partial t} \Psi({\bf R}_{12},t)= -\frac{\hbar^2}{2
M}\frac{\partial^2}{\partial \mathbf{R}_{12}^2}\Psi({\bf R}_{12},t)\cr+\int d {\bf
r}_{12} d {\bf r}_3 d {\bf r}_4 \left[A[{\bf r}_i] \Psi({\bf R}_{12},t)|\Psi( {\bf
R}_{34},t)|^2 + B[{\bf r}_i]\Psi( {\bf R}_{32},t)\Psi^*( {\bf R}_{34},t)\Psi( {\bf
R}_{14},t)\right],
\end{eqnarray}
where $M=m_e+m_h$ is the mass of the pair. The notations $A[{\bf r}_i]$ and $B[{\bf
r}_i]$ stand for the functions of four radius vectors,
\begin{equation}\label{8}
 A[{\bf r}_i]=V_d({\bf r}_1,{\bf r}_2,{\bf r}_3,{\bf r}_4 )|\phi_0( {\bf
r}_{12})|^2 |\phi_0( {\bf r}_{34})|^2,
\end{equation}
\begin{equation}\label{9}
 B[{\bf r}_i]=-V_{ex}({\bf r}_1,{\bf r}_2,{\bf r}_3,{\bf r}_4 )\phi_0^*( {\bf r}_{12})
 \phi_0( {\bf r}_{32})\phi_0^*( {\bf
r}_{34})\phi_0( {\bf r}_{14}),
\end{equation}
where
\begin{equation}\label{10}
  V_d({\bf r}_1,{\bf r}_2,{\bf r}_3,{\bf r}_4
)=V_{ee}({ r}_{13})+V_{hh}({ r}_{24})+V_{eh}({ r}_{14})+V_{eh}({r}_{23}),
\end{equation}
\begin{equation}\label{11}
  V_{ex}({\bf r}_1,{\bf r}_2,{\bf r}_3,{\bf r}_4
)=V_{ee}({ r}_{13})+V_{hh}({ r}_{24})+\frac{1}{2}\left[(V_{eh}({ r}_{12})+V_{eh}({
r}_{34})+V_{eh}({ r}_{14})+V_{eh}({ r}_{23})\right],
\end{equation}
the indices 1 and 3 relate to electrons, and indices 2 and 4 relate to holes. Equation
(\ref{7}) can be interpreted as the generalization of the Gross-Pitaevskii equation for
the BEC of particles with internal degrees of freedom.

We consider the solution of Eq. (\ref{7}) which describes a flowing BEC and takes into
account small fluctuations of the order parameter,
\begin{equation}\label{12}
\Psi({\bf R},t)=e^{-\frac{i {\mu}
t}{\hbar}+i\varphi(\mathbf{R})}\sqrt{n}\left(1+u_{\mathbf{k}} e^{i({\bf k}\cdot{\bf
R}-\omega t)}+v_{\mathbf{k}}^* e^{-i({\bf k}\cdot{\bf R}-\omega t)}\right),
\end{equation}
where $\mu$ is the chemical potential of pairs counted from $E_0$,  $n$ is the density of
the condensate, and $\varphi(\mathbf{R})$ is its phase. The quantities $u_\mathbf{k}$ and
$v_\mathbf{k}$ are the dimensionless amplitudes of the fluctuations.

The gradient of the phase and the flow velocity $\mathbf{V}$ are related by equation
$\mathbf{V}={\hbar}\nabla\varphi/{M}$.
 The chemical potential $\mu$ depends on
 $n$ and $V$,
\begin{equation}\label{13}
    \mu=\gamma n+\frac{M V^2}{2}.
 \end{equation}
The constant $\gamma=\gamma_0^{(d)}+\gamma_0^{(ex)}$ accounts for the direct Coulomb
interaction
\begin{equation}\label{13a}
\gamma_0^{(d)}=\int d {\bf r}_{12} d {\bf r}_3 d {\bf r}_4 A[{\bf r}_i]
\end{equation}
 and  the exchange  Coulomb  interaction,
\begin{equation}\label{13b}
\gamma_0^{(ex)}=\int d {\bf r}_{12} d {\bf r}_3 d {\bf r}_4 B[{\bf r}_i].\end{equation}
 The relation
(\ref{13}) can be obtained from (\ref{7}) and (\ref{12}) at
$u_\mathbf{k}=v_\mathbf{k}=0$.

In the approximation linear in $u_\mathbf{k}$ and $v_\mathbf{k}$,  Eq. (\ref{7}) yields
\begin{equation}\label{14}
    \left(
      \begin{array}{cc}
        \epsilon_k+\hbar \mathbf{k}\mathbf{V}
        +(\gamma_k^{(d)}+\gamma_k^{(ex,1)})n
       & (\gamma_k^{(d)}+\gamma_k^{(ex,2)})n
        \\
       (\gamma_k^{(d)}+\gamma_k^{(ex,2)})n&
        \epsilon_k-\hbar \mathbf{k}\mathbf{V}
        +(\gamma_k^{(d)}+\gamma_k^{(ex,1)})n\\
      \end{array}
    \right)\left(
               \begin{array}{c}
                 u_{\mathbf{k}} \\
                 v_{\mathbf{k}} \\
                               \end{array}
             \right)=\hbar\omega\left(
               \begin{array}{c}
                 u_{\mathbf{k}} \\
                 -v_{\mathbf{k}} \\
               \end{array}
             \right).
\end{equation}
Here, $\epsilon_k=\hbar^2 k^2/2 M$ is the kinetic energy of the pair,
\begin{equation}\label{15}
     \gamma_k^{(d)}=\int d {\bf r}_{12} d {\bf r}_3 d {\bf r}_4  A[{\bf
    r}_i] e^{i {\bf k}\cdot({\bf R}_{34}-{\bf R}_{12})},
\end{equation}
\begin{equation}\label{16}
  \gamma_k^{(ex,1)}=\int d {\bf r}_{12} d {\bf r}_3 d
{\bf r}_4 B[{\bf
    r}_i]\left(e^{i {\bf k}\cdot({\bf R}_{32}-{\bf R}_{12})}+e^{i {\bf k}\cdot({\bf R}_{14}-{\bf
    R}_{12})}-1\right)
\end{equation}
and
\begin{equation}\label{17}
  \gamma_k^{(ex,2)}=\int d {\bf r}_{12} d {\bf r}_3 d {\bf r}_4 B[{\bf
    r}_i]e^{i {\bf k}\cdot({\bf R}_{34}-{\bf R}_{12})}.
\end{equation}
At $k=0$, the quantity (\ref{15}) reduces to $\gamma_0^{(d)}$ (\ref{13a}) and the
quantities (\ref{16}) and (\ref{17}) reduce  to $\gamma_0^{(ex)}$ (\ref{13b}).

From Eq. (\ref{14}), we obtain the spectrum of collective excitations,
\begin{equation}\label{18}
   \hbar \omega(k) =E_0(k)+\hbar \mathbf{k}\mathbf{V},
\end{equation}
where
\begin{equation}\label{19}
   E_0(k)=\sqrt{\left(\epsilon_k+
    [\gamma_{k}^{(ex,1)}-\gamma_{k}^{(ex,2)}]n\right)
    \left(\epsilon_k+
    [2\gamma_k^{(d)}+\gamma_{k}^{(ex,1)}+\gamma_{k}^{(ex,2)}]n\right)}
\end{equation}
is the spectrum of the condensate at rest obtained in \cite{10}.
 The sign before the square root in (\ref{19}) is determined by
  the condition $E_0(k)=\epsilon_k$ at $\gamma_k^{(d)}=\gamma_{k}^{(ex,1(2))}=0$.
As was expected the energies (\ref{18}) and (\ref{19}) are related by the
 Galilean transformation.

To get an approximate analytical expression for the  spectrum (\ref{19}) we consider the
limit of large interlayer distances $d\gg a_0$, where $a_0=\hbar^2\varepsilon/m_* e^2$ is
the effective Bohr radius of the electron-hole pair. In this case, the potential
$V_{e-h}(r)$ in Eq. (\ref{6}) is approximated by its quadratic order expansion near
$r=0$, and the function $\phi_0({\bf r})$ is approximated by the ground-state wave
function of the two-dimensional harmonic oscillator,
\begin{equation}\label{20}
    \phi_0({\bf r})=\frac{1}{\sqrt{\pi}r_0}e^{-\frac{r^2}{2 r_0^2}},
\end{equation}
where $r_0=\sqrt[4]{a_0d^3}$. The parameter $r_0$ determines  the linear size of the
electron-hole pair in the basal plane.

For simplicity, we consider the case $m_e=m_h$. This is  a good approximation for TMD
double layers (the general case $m_e\ne m_h$ is analyzed in Appendix \ref{aa}). For
$m_e=m_h$ and  $\phi_0({\bf r})$ given by Eq. (\ref{20}), the functions
(\ref{15})-(\ref{17}) are given by the expression \cite{10}:
\begin{eqnarray}\label{21}
 \gamma_k^{(d)}=\frac{4\pi e^2}{\varepsilon k}(1-e^{- k
 d})e^{-\frac{k^2 r_0^2}{8}},
\end{eqnarray}
\begin{eqnarray}\label{22}
 \gamma_k^{(ex,1)}=-\frac{4\pi e^2 r_0}{\varepsilon }
\Bigg[\sqrt{\frac{\pi}{2}}\left(\mathrm{fa}\left(\frac{k^2
r_0^2}{16}\right)+e^{-\frac{k^2 r_0^2}{8}}-1\right)-2 e^{-\frac{k^2 r_0^2}{8}}
\mathrm{fd}\left(\frac{k r_0}{4} \right) +\mathrm{fd}\left(0\right)\Bigg],
\end{eqnarray}
\begin{eqnarray}\label{23}
 \gamma_k^{(ex,2)}=-\frac{4\pi e^2 r_0}{\varepsilon }
\Bigg[\sqrt{\frac{\pi}{2}}e^{-\frac{ k^2 r_0^2}{8}}\mathrm{fa}\left(\frac{k^2
r_0^2}{16}\right)-e^{-\frac{k^2 r_0^2}{4}}\frac{\mathrm{fd}\left(\frac{k r_0}{2}
\right)+\mathrm{fd}\left(0\right)}{2} \Bigg].
\end{eqnarray}
In (\ref{22}) and (\ref{23}), we use the shorthand notations
\begin{equation}\label{24}
    \mathrm{fa}(y)=e^{-y}I_0(y),
\end{equation}
\begin{equation}\label{25}
    \mathrm{fd}(k)=\int_0^\infty e^{-\frac{3p^2}{8}-p
d/r_0}I_0\left(p k \right) d p,
\end{equation}
where  $I_0(x)$ is the modified Bessel function.

\begin{figure}
\begin{center}
\includegraphics[width=6cm]{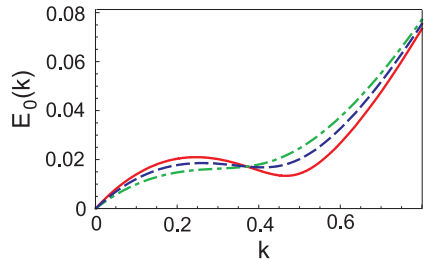}
\end{center}
\caption{Collective mode spectrum calculated for $d=5 a_0$ and $n=4\cdot 10^{-3}
a_0^{-2}$ (solid curve), $n=3\cdot 10^{-3} a_0^{-2}$ (dashed curve), and $n=2\cdot
10^{-3} a_0^{-2} $ (dash-dotted curve). The energy $E_0(k)$ is given in units of
$e^2/\varepsilon a_0=2 Ry_{eff}$ (doubled effective Rydberg),  and $k$ in units of
$a_0^{-1}$.}\label{f1}
\end{figure}

Below we use the effective Bohr radius $a_0$ as the unit of length.   In these units the
density $n$ is measured in $a_0^{-2}$, and the wave vector $k$ is measured in $a_0^{-1}$.
For TMD embedded in hexagonal  boron nitride, this length is about $a_0\approx 1$ nm and
$a_0^{-2}\approx 10^{14}$ cm$^{-2}$. The dispersion curves $E_0(k)$ calculated at three
different densities, i.e., $n=2\cdot 10^{-3}a_0^{-2}$, $n=3\cdot 10^{-3}a_0^{-2}$, and
$n=4\cdot 10^{-3}a_0^{-2}$  and $d=5 a_0$ are shown in Fig. \ref{f1}. These parameters
correspond to the low-density limit $n r_0^2\ll 1$. The superfluid-supersolid transition
is the first-order transition \cite{my2}. The
 transition is accompanied by a jump in the density $n$.
At $d=5 a_0$ and $m_e=m_h$, the density increases from $n\approx 4.1\cdot 10^{-3 }
a_0^{-2}$ to $n\approx 4.3\cdot 10^{-3}a_0^{-2}$ at the transition point. Thus, all
curves in Fig. \ref{f1} correspond to the stable uniform phase. One can see that the
roton-type minimum emerges in a certain range of parameters. Another feature of the
spectrum in Fig. \ref{f1} is that the difference  of the group and phase velocities as a
function of $k$ changes its sign (in contrast to Bose gases with contact interaction
where this difference is positive for all $k$).  These two features reveal themselves in
an essential modification of the stationary wave pattern as compared to the wave pattern
in the atomic BEC \cite{3, 3-1, 3-2}.

In what follows, we calculate the stationary wave patterns for the parameters that
correspond to the system far from the superfluid-supersolid transition line (the
dash-dotted curve in Fig. \ref{f1}) and to the system close the transition line (solid
curve in Fig. \ref{f1}).

\section{Kinematic approach} \label{s3}

To obtain the wave crest pattern generated by a point obstacle, we use the kinematic
approach \cite{1}.  We consider the two-dimensional problem in the rest frame with the
obstacle at the origin and the flow velocity vector $\mathbf{V}=(-V,0)$.
 In this frame of reference the stationary wave crests do not move. The condition of that is
\begin{equation}\label{26}
 \hbar \omega(k) = E_0(k)-\hbar k V \cos\xi=0,
\end{equation}
where $\xi$ is the angle between the wave vector $\mathbf{k}$ and the $x$ axis. Equation
(\ref{26}) implicitly defines  the functions $k(\xi)$ and $k_x(k_y)$ ($k_x$, $k_y$ are
the components of $\mathbf{k}$ and $k$ is its modulus). In the general case, these
functions are multiple valued. The number of branches is equal to the number of
intersections of the dispersion curve, $\omega=E_0(k)$, with the line $\omega=V k \cos
\xi$.

The phase of the stationary wave is given by the integral
\begin{equation}\label{26-1}
    \theta(\bf{r})=\int_0^{\bf{r}}\mathbf{k}(\mathbf{r}')d \mathbf{r}'.
\end{equation}
It follows from (\ref{26-1}) that the components of $\mathbf{k}$ satisfy the relation
\begin{equation}\label{26-2}
    \frac{\partial k_y}{\partial x}=\frac{\partial k_x}{\partial y}.
   \end{equation}
The right-hand side of Eq. (\ref{26-2}) can be expressed through the derivative of the
function $k_x(k_y)$:  ${\partial k_x}/{\partial y}=({d k_x}/{d k_y})({\partial
k_y}/{\partial y})$. Substituting the function $k_x(k_y)$ into (\ref{26}) we obtain the
identity
\begin{equation}\label{26-5}
   E_0\left(\sqrt{[k_x(k_y)]^2+k_y^2}\right)-\hbar V k_x(k_y)\equiv 0.
\end{equation}
The derivative of the left-hand side of Eq. (\ref{26-5}) with respect to $k_y$ is equal
to zero. It yields
\begin{equation}\label{26-6}
 \frac{d k_x}{d k_y}
 =-\frac{\frac{d E_0}{d k}\frac{k_y}{k}}{\frac{d E_0}{d k}\frac{k_x}{k}-\hbar V }.
\end{equation}
Equation (\ref{26-6}) can be rewritten as
\begin{equation}\label{26-3}
   \frac{d k_x}{d k_y}=-\frac{{v}_{g,y}}{{v}_{g,x}},
\end{equation}
where ${v}_{g,x}$ and ${v}_{g,y}$ are the components of
\begin{equation}\label{27}
    \mathbf{v}_g(\mathbf{k})=\frac{d \omega(\mathbf{k})}{d \mathbf{k}},
\end{equation}
the group velocity of collective excitations in the rest frame.

It follows from Eqs. (\ref{26-2}) and (\ref{26-3}) that the components of the vector
$\mathbf{k}$ are the constants along the line
\begin{equation}\label{28-1}
    \frac{x}{y}=\frac{{v}_{g,x}}{{v}_{g,y}}.
\end{equation}
This line is along the  group velocity (\ref{27}). Thus the group velocity
$\mathbf{v}_g(\mathbf{k})$ determines the direction of propagation of the stationary
wave, with the wave vector $\mathbf{k}$,  in the rest frame of reference. The integral
(\ref{26-1}) calculated along the line $\mathbf{k}=const$ is equal to
\begin{equation}\label{29} \theta(r,\xi)=k r \cos(\xi-\chi),
\end{equation}
where  $\chi$ is the angle between the group velocity (\ref{27}) and the $x$ axis. The
angle $\chi$ is the function of $\xi$. The dependence $\chi(\xi)$ can be found from the
expressions for the components of $\mathbf{v}_g$ at $k=k(\xi)$:
\begin{equation}\label{28}
    {v}_{g,x}(\xi)=\frac{1}{\hbar}\frac{d E_0(k)}{d k}\Big|_{k=k(\xi)}\cos\xi
    -V,\quad {v}_{g,y}(\xi)
    =\frac{1}{\hbar}\frac{d E_0(k)}{d k}\Big|_{k=k(\xi)}\sin\xi.
\end{equation}

At the crest lines, the phase (\ref{29}) is a multiply of $2\pi$,
\begin{equation}\label{30}
\theta_N(\xi)= 2\pi N \mathrm{sign}[\cos[\xi-\chi(\xi)]],
\end{equation}
 where $N=1,2,3,\ldots$. The function $\mathrm{sign}(x)$ in (\ref{30}) takes into account that
 the sign of $\theta$ in Eq.(\ref{29}) coincides with the sign of $\cos(\xi-\xi)$ (
 by definition, $k$ and $r$ are the positive quantities).

From Eq. (\ref{29}), we obtain the parametric equations that determine the crest lines
\begin{eqnarray} \label{31}
  x_{N}(\xi) &=& \frac{\theta_{N}(\xi) \cos[\chi(\xi)]}
  {k(\xi)\cos[\xi-\chi(\xi)]},\cr
  y_{N}(\xi)&=& \frac{\theta_{N}(\xi) \sin[\chi(\xi)]}{k(\xi)\cos[\xi-\chi(\xi)]}.
\end{eqnarray}

Taking into account the relations (\ref{26}) and (\ref{28}) we present Eqs. (\ref{31}) in
the form
\begin{eqnarray} \label{32}
  x_N(\xi) &=&2\pi N\frac{ v_{g0}(\xi)\cos\xi- V}
  {k(\xi)\left[v_{g0}(\xi)-v_{p0}(\xi)\right]}\mathrm{sign}\left[v_{g0}(\xi)-v_{p0}(\xi)\right],
  \cr
  y_N(\xi)&=& 2\pi N\frac{ v_{g0}(\xi)\sin \xi}
  {k(\xi)\left[v_{g0}(\xi)-v_{p0}(\xi)\right]}
  \mathrm{sign}\left[v_{g0}(\xi)-v_{p0}(\xi)\right],
\end{eqnarray}
where $v_{g0}(\xi)=(d E_0(k)/d k)|_{k=k(\xi)}$ and $v_{p0}(\xi)=E_0(k)/k|_{k=k(\xi)}$,
respectively, are the  group and the phase velocities of the collective mode in the
condensate at rest (negative $v_{g0}$ corresponds to oppositely directed
$\mathbf{v}_{g0}$ and $\mathbf{k}$). The functions $v_{g0}(\xi)$, $v_{p0}(\xi)$,
$x_N(\xi)$ and $y_N(\xi)$ have the same number of branches as the function $k(\xi)$ has.
For the dispersion curves shown in Fig. \ref{f1} the function $k(\xi)$ has two branches,
the lower and the upper ones. The lower branch corresponds to $k$ for which
$v_{g0}<v_{p0}$, and the upper branch corresponds to $k$ for which  $v_{g0}>v_{p0}$.

One can see from (\ref{32}) that the intersections of the crest lines with the $x$ axis
correspond to $\xi=0$ or to $v_{g0}(\xi)=0$. For the upper branch of $k(\xi)$,  the
velocity $v_{g0}\ne 0$ for all $\xi$ and the $N$-th crest line intersects the $x$ axis
only once [at $x=x_N(0)$]. There can be up to three intersections for the crest line
associated with the lower branch of $k(\xi)$. The number of intersections depends on the
presence (absence) of the roton-type minimum and on the value of $V$.   At
$v_{g0}(\xi)=v_{p0}(\xi)$ the coordinates $x_N(\xi)$ and $y_N(\xi)$ approach infinity.
The condition $v_{g0}=v_{p0}$ is reached at $k\to 0$ and at the minimum of the phase
velocity.

\begin{figure}
\begin{center}
\includegraphics[width=6cm]{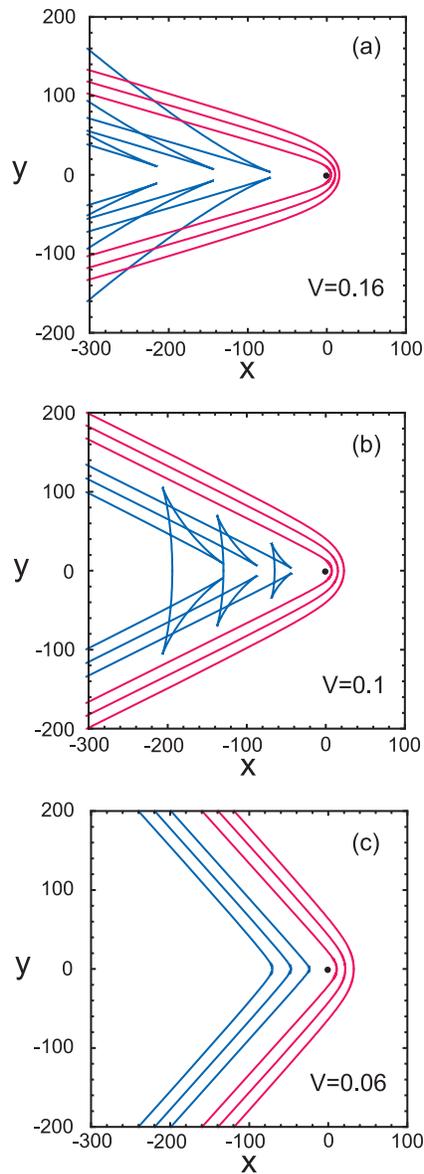}
\end{center}
\caption{Crests of stationary waves induced by a point obstacle at  $n=2\cdot 10^{-3}
a_0^{-2}$ and $d=5 a_0$.  The flow velocity  is equal to (a) $V=0.16$, (b) $V=0.1$, and
(c) $V=0.06$ in $\alpha c/\varepsilon$ units. Red (blue) crests correspond to the
collective modes with $v_{g0}<v_{p0}$ ($v_{g0}>v_{p0}$). Only the first three crests are
shown. The black dot indicates the obstacle location. The spatial scale unit is
$a_0$.}\label{f2}
\end{figure}

To illustrate the complexity of the wave crest pattern we consider the system far from
the superfluid-supersolid transition line and the system close to the transition line.
The spectrum of the system far from the transition line (dash-dotted curve in Fig.
\ref{f1}) has no roton-type minimum. One can distinguish two special phase velocities:
$v_0$, the sound velocity at $k\to 0$, and $v_{min}$, the minimum phase velocity.  The
convenient units for the velocities are $2 Ry_{eff} a_0/\hbar=\alpha c/\varepsilon$,
where $c$ is the speed of light, and $\alpha\approx 1/137$ is the fine structure
constant. In these units $v_0\approx 0.128$ and $v_{min}\approx 0.045$ at $n=2\cdot
10^{-3}  a_0^{-2}$ and $d=5 a_0$. In Fig. \ref{f2}, the wave crest patterns calculated
for $V=0.16$, $V=0.1$, and $V=0.06$ are presented. One can see that the stationary wave
patterns at $V>v_0$ and  $v_0>V>v_{min}$ differ from each other. In the former case
stationary waves are not excited in a narrow wedge behind the obstacle. In the latter
case stationary waves fill all the space. At $V<v_{min}$, stationary waves are not
excited. Within the range $v_0>V>v_{min}$, the pattern is changed qualitatively under
variation in the flow velocity: at larger $V$, the wave crests have cusps and
self-intersections, and at smaller $V$, the cusps and intersections disappear.

\begin{figure}
\begin{center}
\includegraphics[width=12cm]{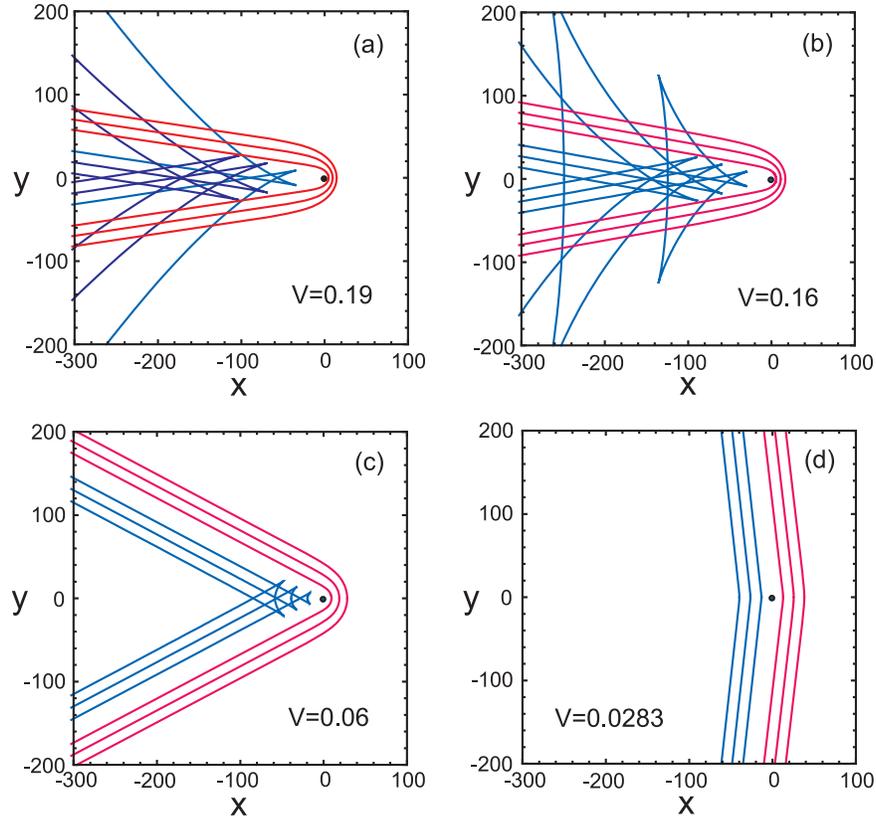}
\end{center}
\caption{The same as in Fig. \ref{f2} for $n=4\cdot 10^{-3} a_0^{-2}$, $d=5 a_0$, and the
flow velocity (a) $V=0.19$, (b) $V=0.16$, (c) $V=0.06$,  and (d) $V=0.0283$   (in $\alpha
c/\varepsilon$ units).}\label{f3}
\end{figure}

One can distinguish four special velocities  for the system close to the
superfluid-supersolid transition line. Two of them, $v_0$ and $v_{min}$, are defined
above. Two other velocities, $v_1$ and $v_2$, are the phase velocities at the local
maximum and at the local minimum in the dispersion curve $E_0(k)$.   For $n=4\cdot
10^{-3}a_0^{-2}$ and $d=5 a_0$, the special  velocities are $v_0\approx 0.18$,
$v_1=0.085$, $v_2=0.0287$ and $v_{min}=0.0281$ (in $\alpha c/\varepsilon$ units). In Fig.
\ref{f3}, we present the crest line patterns for $V=0.19$ ($V>v_0$), $V=0.16$
($v_0>V>v_1$), $V=0.06$ ($v_1>V>v_2$), and $V=0.0285$ ($v_2>V>v_{min}$). One can see that
in this case, the wave crest pattern is also modified significantly under variation in
the flow velocity.

 In all cases, the wave crest pattern contains a family of crests located outside the
 wedge-shaped region with the semiangle $\theta=\arccos(v_{min}/V)$. A similar family
 is observed in Bose gases with contact interaction \cite{e2,e3,3, 3-1, 3-2}.
 The other families of crests (absent in the atomic BEC) are located entirely behind
 the obstacle. In the general case, they contain cusps and intersections.

\section{Dynamical approach} \label{s4}

 For the comparison
of the theory with the experiment, it is instructive to find relative amplitudes of
stationary waves. It can be done in the framework of the dynamical approach \cite{1}.

To apply the dynamical approach to the BEC of spatially indirect excitons, we take into
account the interaction with the obstacle in Eq. (\ref{7}),
\begin{eqnarray}\label{33}
i \hbar \frac{\partial}{\partial t} \Psi({\bf R}_{12},t)= -\frac{\hbar^2}{2
M}\frac{\partial^2}{\partial\mathbf{R}_{12}^2}\Psi({\bf R}_{12},t)+
U(\mathbf{R}_{12})\Psi({\bf R}_{12},t)\cr+\int d {\bf r}_{12} d {\bf r}_3 d {\bf r}_4
\left[A[{\bf r}_i] \Psi({\bf R}_{12},t)|\Psi( {\bf R}_{34},t)|^2 + B[{\bf r}_i]\Psi( {\bf
R}_{32},t)\Psi^*( {\bf R}_{34},t)\Psi( {\bf R}_{14},t)\right].
\end{eqnarray}
 The potential $U(\mathbf{R})$ describes the interaction of an electron-hole pair,
  which center of mass has the coordinate $\mathbf{R}$, with the obstacle
  at the origin.
  The solution of (\ref{33}) is sought in the form
\begin{equation}\label{34}
   \Psi({\bf R},t)=\Psi_0({\bf R},t)+\Psi_1({\bf R},t)=e^{-i\frac{\mu}{\hbar}t}e^{i \mathbf{R}\nabla
   \varphi}\left[\sqrt{n}+\tilde{\Psi}_1({\bf R},t)\right].
\end{equation}
The first term in the square brackets in (\ref{34}) corresponds to the uniform
condensate. The term $\Psi_1({\bf R},t)$ is caused by the interaction with the obstacle.
The point obstacle is modeled by the potential
\begin{eqnarray} \label{35}
     U(\mathbf{R},t) = U_0 \delta(\mathbf{R})e^{\eta t},
\end{eqnarray}
where $\eta=+0$. We imply that the interaction with the obstacle was switched on at $t\to
-\infty$ and  look for a response at $t=0$.  The response is assumed to be small, linear
in $U_0$. The  divergence of (\ref{35}) at $t\to \infty$ is not relevant because due to
the casuality principle the response of the system to the obstacle at $t=t_0$ depends on
$U(\mathbf{R},t')$ with $t'\leq t_0$. The time dependence of $\tilde{\Psi}_1({\bf R},t)$
has the form $\tilde{\Psi}_1({\bf R},t)=\tilde{\Psi}_1({\bf R})e^{\eta t}$. In the linear
in $U_0$ order we obtain the following equation for $\tilde{\Psi}_1(\mathbf{R})$:
\begin{eqnarray} \label{36}
-\frac{\hbar^2}{2 M}\frac{\partial^2}{\partial\mathbf{R}_{12}^2}\tilde{\Psi}_1({\bf
R}_{12})-i\hbar \mathbf{V} \cdot
\frac{\partial}{\partial\mathbf{R}_{12}}\tilde{\Psi}_1({\bf R}_{12})- i\hbar\eta
\tilde{\Psi}_1({\bf R}_{12})\cr +\int d {\bf r}_{12} d {\bf r}_3 d {\bf r}_4 A[{\bf
r}_i]{n}\left[ \tilde{\Psi}_1({\bf R}_{34})+\tilde{\Psi}^*_1({\bf R}_{34})\right]\cr
+\int d {\bf r}_{12} d {\bf r}_3 d {\bf r}_4 B[{\bf r}_i]{n}\left[ \tilde{\Psi}_1({\bf
R}_{32})+\tilde{\Psi}_1({\bf R}_{14})+\tilde{\Psi}^*_1({\bf R}_{34})-\tilde{\Psi}_1({\bf
R}_{12})\right]=-\sqrt{n}U_0\delta(\mathbf{R}_{12}).
\end{eqnarray}
From (\ref{36}), we find the system of equations for the Fourier components of
$\tilde{\Psi}_1(\mathbf{R})$:
\begin{eqnarray}\label{37}
\left[\epsilon_k+\hbar(\mathbf{k}\mathbf{V}-i\eta)+(\gamma_k^{(d)}+\gamma_k^{(ex,1)})n\right]
\tilde{\Psi}_1({\bf k})+(\gamma_k^{(d)}+\gamma_k^{(ex,2)})n\tilde{\Psi}^*_1({-\bf
k})=-\sqrt{n}U_0, \cr (\gamma_k^{(d)}+\gamma_k^{(ex,2)})n\tilde{\Psi}_1({\bf
k})+\left[\epsilon_k-\hbar(\mathbf{k}\mathbf{V}-i\eta)+(\gamma_k^{(d)}+\gamma_k^{(ex,1)})n\right]
\tilde{\Psi}^*_1(-{\bf k})=-\sqrt{n}U_0.
\end{eqnarray}
In obtaining (\ref{37}) we take into account that  $\gamma_k^{(d)}$ and
$\gamma_k^{(ex,i)}$ depend on the modulus $\mathbf{k}$. The solution of the system
(\ref{37}) is
\begin{equation}\label{38}
   \tilde{\Psi}_1({\bf k})=-\frac{\epsilon_k+(\gamma_k^{(ex,1)}-\gamma_k^{(ex,2)})n
   -\hbar(\mathbf{k}\mathbf{V}-i\eta)}{[E_0(k)]^2-\hbar^2(\mathbf{k}\mathbf{V}-i\eta)^2}\sqrt{n}U_0.
\end{equation}

In the same order in $U_0$ the spatial variation of the density is given by the equation
\begin{equation}\label{39}
    \delta n(\mathbf{r})=\sqrt{n}\left[\tilde{\Psi}_1({\bf r})+\tilde{\Psi}_1^*({\bf r})\right]=
\frac{\sqrt{n}}{(2\pi)^2}\int d^2 k  e^{i \mathbf{k}\mathbf{r}}\left[\tilde{\Psi}_1({\bf
k})+\tilde{\Psi}_1^*(-{\bf k})\right].
\end{equation}
Substitution of (\ref{38}) into (\ref{39}) yields
\begin{equation}\label{40}
   \delta  n(\mathbf{r})=
-\frac{n U_0}{2\pi^2}\int d \mathbf{k}  e^{i
\mathbf{k}\mathbf{r}}\frac{\epsilon_k+(\gamma_k^{(ex,1)}-\gamma_k^{(ex,2)})n
   }{[E_0(k)]^2-\hbar^2(\mathbf{k}\mathbf{V}-i\eta)^2}.
\end{equation}
Introducing the polar angle $\chi$ [$\mathbf{r}=(r\cos\chi,r\sin\chi)$] and considering
$\mathbf{V}=(-V,0)$, we get
\begin{equation}\label{42}
  \delta   n(\mathbf{r})=
-\frac{n U_0}{\pi^2}\mathrm{Re}\int_{-\frac{\pi}{2}}^{\frac{\pi}{2}} d\xi \int_0^\infty d
k e^{i k r
\cos(\xi-\chi)}\frac{k\left[\epsilon_k+(\gamma_k^{(ex,1)}-\gamma_k^{(ex,2)})n\right]
   }{[E_0(k)]^2-\hbar^2(k V \cos\xi+i\eta)^2}.
\end{equation}
In obtaining (\ref{42}), we take into account that the integrand in Eq.(\ref{42})
transforms into the complex conjugated expression under substitution, $\xi\to \xi+\pi$.

We evaluate the integral over $k$ in Eq. (\ref{42}) using the residue theorem. Then we
calculate the integral over $\xi$ in the stationary phase approximation.  Details are
given in  Appendix \ref{bb}. The answer has the form
\begin{equation}\label{48}
  \delta   n(\mathbf{r})={n U_0}
\sqrt{\frac{2}{\pi}}\sum_{\lambda,i} \left[\frac{\sin\left[ f_\lambda(\xi)
+\frac{\pi}{4}\mathrm{sign}[f''_\lambda(\xi)]\right]} {\sqrt{|f''_\lambda(\xi)|}}
s_\lambda[\xi,\chi]\right]
\left[\frac{k\left[\epsilon_k+(\gamma_k^{(ex,1)}-\gamma_k^{(ex,2)})n\right]
   }{E_0(k)\left[\frac{d E_0(k)}{d
k}-\frac{E_0(k)}{k}\right]}\right]\Bigg|_{\begin{array}{c}
                                            k=k_\lambda(\xi), \\
                                            \xi=\xi_{\lambda,
i}(\chi)
                                          \end{array}},
                                         \end{equation}
where  the index $\lambda$ is the branch number of the multivalued function $k(\xi)$
defined in the previous section,
\begin{equation}\label{47}
f_\lambda(\xi) = k_\lambda(\xi) r \cos(\chi-\xi)
\end{equation}
is the phase of the exponent in (\ref{42}), and $\xi_\lambda(\chi)$ is the function
reciprocal to the function $\chi_\lambda(\xi)$ introduced in the previous section. In the
general case, the function $\xi_\lambda(\chi)$ at given $\lambda$ is multivalued  and the
index $i$ is the branch number of this function at fixed $\lambda$. The factor
$s_\lambda[\xi,\chi]$ in Eq. (\ref{48}) is defined as
\begin{equation}\label{45}
    s_\lambda(\xi,\chi)=\left\{
      \begin{array}{ll}
        1, &\hbox{ if $\left(\frac{d E_0(k)}{d
k}-\frac{E_0(k)}{k}\right)\Big|_{k=k_\lambda(\xi)}<0$ and $\cos(\xi-\chi)<0$;} \\
        -1, &\hbox{ if $\left(\frac{d E_0(k)}{d
k}-\frac{E_0(k)}{k}\right)\Big|_{k=k_\lambda(\xi)}>0$ and $\cos(\xi-\chi)>0$;} \\
        0, & \hbox{otherwise.}
      \end{array}
    \right.
\end{equation}
The amplitude of spatial oscillations of $\delta n(\mathbf{r})$  (\ref{48}) is
proportional to $1/\sqrt{r}$, where $r$ is the distance from the obstacle. The
contribution proportional to $1/r^2$ is neglected in (\ref{48}) (see  Appendix \ref{bb}).

In Fig. \ref{f4}, we present the density plot of $\delta n(\mathbf{r})$ (the stationary
wave pattern) calculated for the same parameters as in Fig. \ref{f2}. One can see that
some  wave crests shown in Figs. \ref{f2}(a) and \ref{f2}(b), are invisible in the
density plots Figs. \ref{f4}(a) and \ref{f4}(b). These wave crests correspond to modes
with small $k$. Indeed, summands in Eq. (\ref{48}) are proportional to $k^{3/2}$ at small
$k$ and, due to this, the relative amplitudes of stationary waves that correspond to
long-wave modes are small.

\begin{figure}
\begin{center}
\includegraphics[width=16cm]{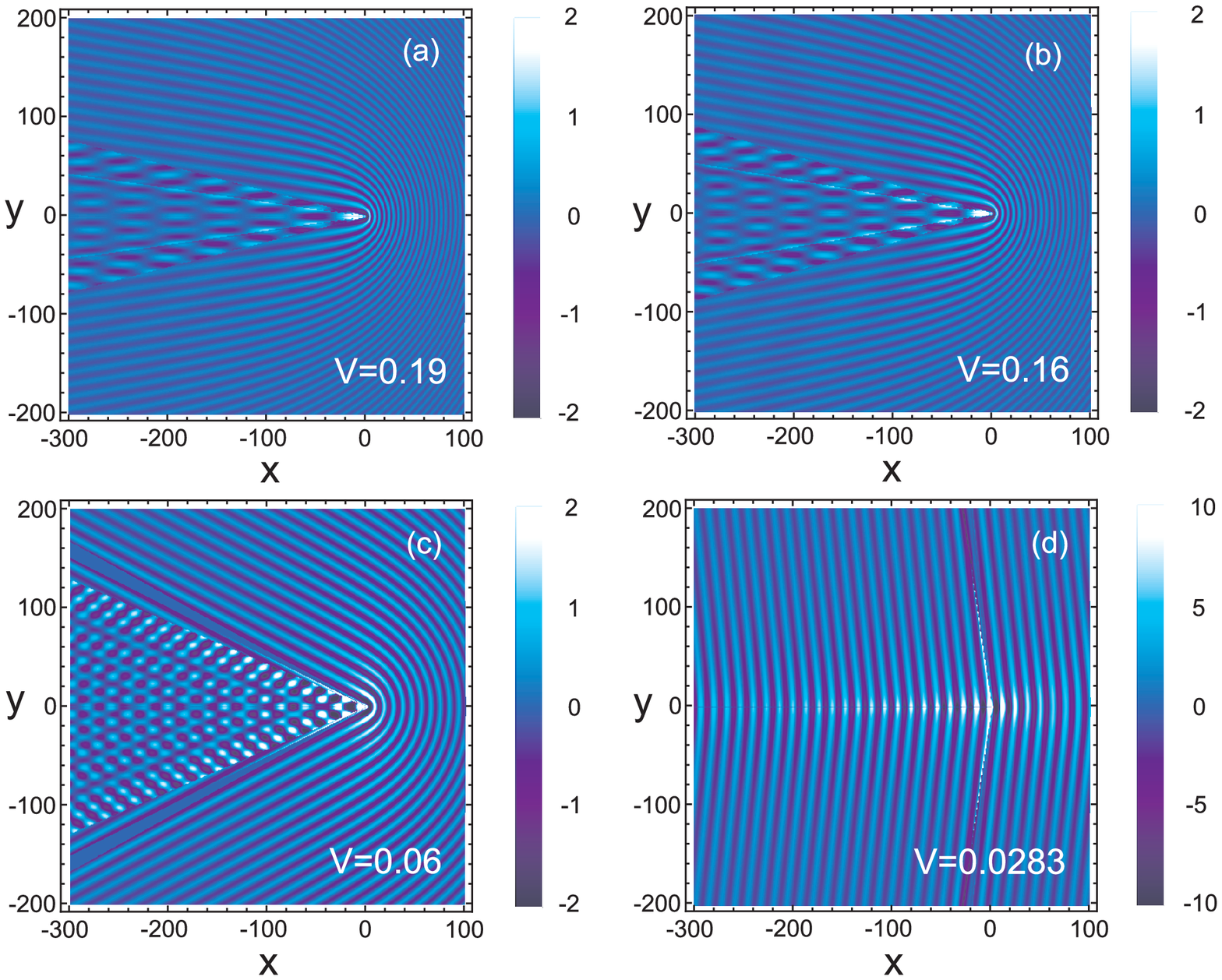}
\end{center}
\caption{Stationary wave pattern induced by the point obstacle in the system close to the
superfluid-supersolid transition line ($n=4\cdot 10^{-3} a_0^{-2}$ and $d=5 a_0$). The
flow velocities $V$ are given in $c/137 \varepsilon$ units. The density variation $\delta
n$ is given in units of $n U_0/(2Ry_{eff} a_0)$.}\label{f5}
\end{figure}

In Fig. \ref{f5}, we present the stationary wave pattern calculated for the same
parameters as in Fig. \ref{f3}. Again we see that stationary waves with small $k$ are
practically invisible in the density plots.

The changes in the stationary wave pattern under variation in the flow velocity are less
spectacular than the changes in the wave crest pattern. Nevertheless, the stationary wave
pattern differs significantly from the one obtained for the atomic BEC \cite{3, 3-1,3-2}
and it is modified qualitatively with a change in the flow velocity and in the density of
the condensate.

\begin{figure}
\begin{center}
\includegraphics[width=8cm]{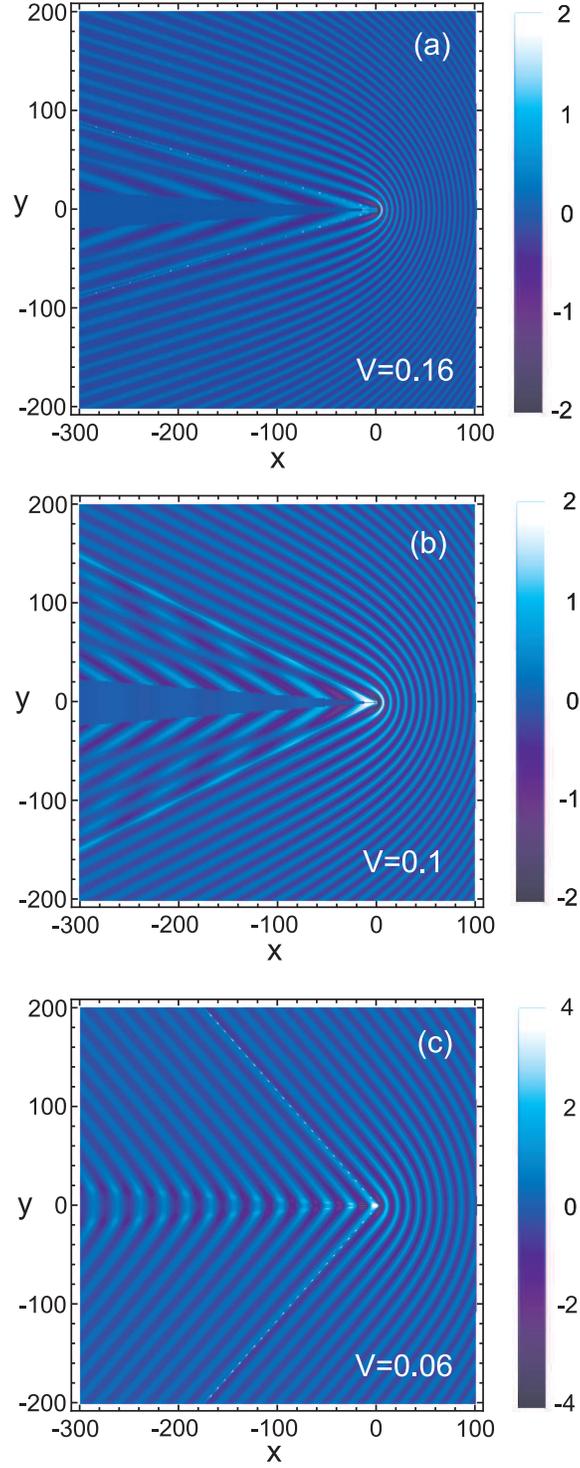}
\end{center}
\caption{ Stationary wave pattern induced by the point obstacle in the system far from
the superfluid-supersolid transition line ($n=2\cdot 10^{-3} a_0^{-2}$ and $d=5 a_0$).
The flow velocities $V$ are given in $c/137 \varepsilon$ units. The density variation
$\delta n$ is given in units of $n U_0/(2Ry_{eff} a_0)$. }\label{f4}
\end{figure}

To visualize stationary waves caused by the long-wave part of the spectrum, one can use a
weak extended obstacle instead of the point one. To illustrate this possibility, we
consider the obstacle in which the interaction with  electron-hole pairs is described by
the Gaussian potential
\begin{equation}\label{63}
    U(\mathbf{R})=\frac{U_0}{2\pi a^2}\exp\left(-\frac{R^2}{2 a^2}\right).
\end{equation}
The potential (\ref{63}) is normalized by the same condition as the potential (\ref{35}):
\begin{equation}\label{65}
    \int d^2 r U(\mathbf{r})=U_0.
\end{equation}
The Fourier components of the potential (\ref{63}) are equal to
\begin{equation}\label{64}
U(\mathbf{k})=U(k)=U_0 \exp\left(-\frac{k^2 a^2}{2}\right).
\end{equation}
For the potential (\ref{63}), the dynamical approach yields
 \begin{equation}\label{62}
   \delta n(\mathbf{r})={n}
\sqrt{\frac{2}{\pi}}\sum_{\lambda,i} U(k) \frac{\sin\left[ f_\lambda(\xi)
+\frac{\pi}{4}\mathrm{sign}[f''_\lambda(\xi)]\right]} {\sqrt{|f''_\lambda(\xi)|}}
s_\lambda[\xi,\chi]\left[\frac{k\left[\epsilon_k+(\gamma_k^{(ex,1)}-\gamma_k^{(ex,2)})n\right]
   }{E_0(k)\left[\frac{E_0(k)}{d
k}-\frac{E_0(k)}{k}\right]}\right]\Bigg|_{\begin{array}{c}
                                            k=k_\lambda(\xi), \\
                                            \xi=\xi_{\lambda,
i}(\chi)
                                          \end{array}}
\end{equation}
Equation (\ref{62}) differs from Eq. (\ref{48}) by the factor $U(k)$.

\begin{figure}
\begin{center}
\includegraphics[width=8cm]{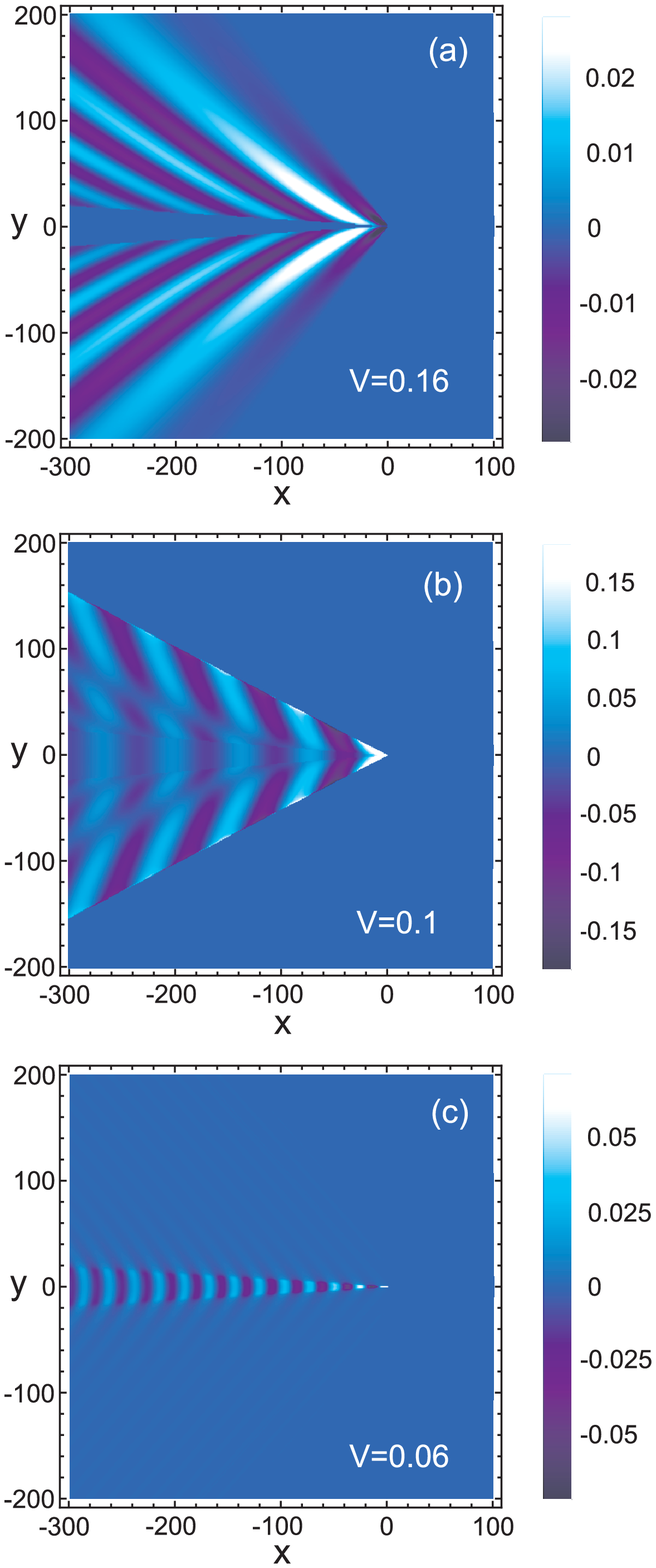}
\end{center}
\caption{The stationary wave patterns induced by a weak extended obstacle for $n=2\cdot
10^{-3} a_0^{-2}$ and $d=5 a_0$. The flow velocities are the same as in Fig. \ref{f4}.
The interaction of the obstacle with electron-hole pairs is described by the Gaussian
potential (\ref{63}) with $a=10 a_0$.}\label{f6}
\end{figure}

\begin{figure}
\begin{center}
\includegraphics[width=16cm]{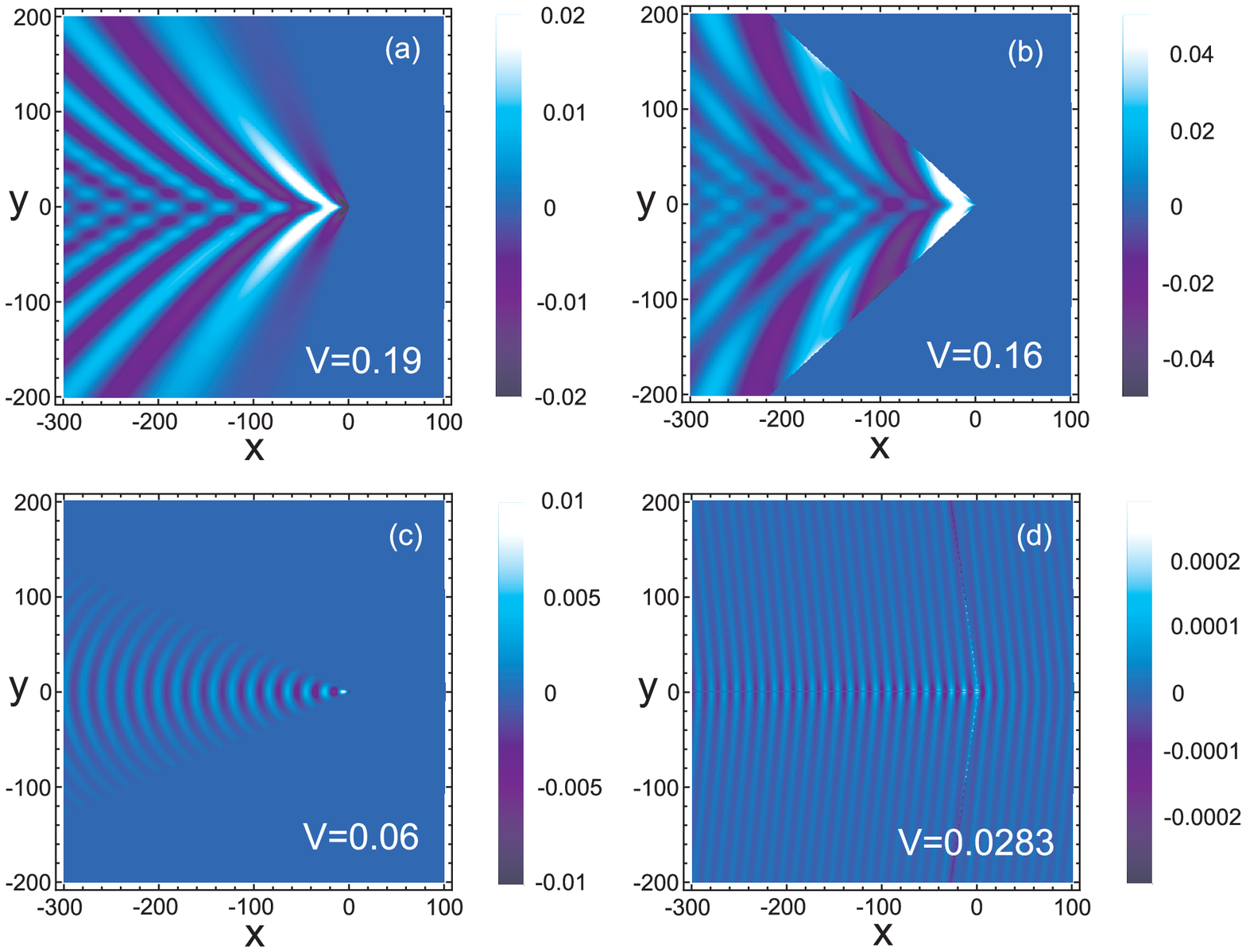}
\end{center}
\caption{The stationary wave patterns induced by a weak extended obstacle for $n=4\cdot
10^{-3} a_0^{-2}$ and $d=5 a_0$. The flow velocities are the same as in Fig. \ref{f5}.
The interaction of the obstacle with electron-hole pairs is described by the Gaussian
potential (\ref{63}) with $a=10 a_0$.}\label{f7}
\end{figure}

We specify the case  $a\gg a_0$. In this case, the factor $U(k)$ suppresses  the
short-wave contribution into $\delta n(\mathbf{r})$ and the long-wave contribution
becomes visible. The stationary wave patterns calculated for the potential (\ref{63})
with $a=10 a_0$  are shown in Figs. \ref{f6} and \ref{f7}. One can see that the wave
crests hidden in Figs. \ref{f4} and \ref{f5} are easily identified in Figs. \ref{f6} and
\ref{f7}, and vice versa. The patterns in Figs. \ref{f6}(b) and \ref{f7}(b) resemble ship
waves  on a surface of deep water.  The waves in Figs. \ref{f6}(b) and \ref{f7}(b)  fill
the wedge-shaped region behind the obstacle, but the semiangle of this wedge is not a
universal quantity. As in the previous case, the stationary wave pattern  is modified
qualitatively under variation in the exciton density and under variation in the flow
velocity.

The spatial resolution should not be less than  $(20 - 40)a_0$ to observe the stationary
wave patterns shown in Figs. \ref{f6}(a), \ref{f6}(b) and  in Figs. \ref{f7}(a),
\ref{f7}(b). In the case of GaAs heterostructures, it corresponds to 200-400 nm. That is
the maximum resolution  achieved in photoluminescence experiments with spatially indirect
excitons \cite{plex}. At lower resolutions, one can also observe stationary waves, but
only at a large distance from the obstacle.  In TMD double layer systems, the effective
Bohr radius is ten times smaller then in  GaAs and the resolution that can be achieved in
photoluminescence experiments is too low to observe the stationary wave pattern in such
structures. Much better spatial resolution (about 10 nm) was achieved in \cite{eel} by
the method of probing of excitons in \cite{eel} based on  obtaining the low-energy loss
spectra using the scanning transmission electron microscope. We may also suggest to
register stationary waves by measuring the spatial variation of the electrostatic
potential near the bilayer system. The most effective method of measuring the local
electrostatic potential uses the single-electron transistor (SET) technique \cite{set}.
This method provides spatial resolution  of the order of the SET size that can be about
100 nm. We also suggest to use the stationary wave pattern [especially the pattern shown
Fig. \ref{5}(d)] as a grating for a generation of short-wave plasmons in a nearby
conducting layer.

\section{Conclusion}

In conclusion we have considered  stationary waves occurring in the BEC of spatially
indirect excitons in bilayers. Under an increase in the exciton density or in the
distance between the layers the uniform state of the exciton condensate becomes unstable
with respect to the transition to the supersolid state. The roton-type minimum in the
collective mode spectrum emerges as a precursor of such a transition. The presence of
this minimum reveals itself in a number of features in the stationary wave pattern. The
observation of such features and qualitative changes in the stationary wave pattern under
an increase in the exciton density can be a bright manifestation of superfluidity of the
exciton gas and its proximity to the superfluid-supersolid transition line.

Two possible systems where the effects predicted in this paper can be observed are TMD
bilayers and double quantum well GaAs heterostructures. The advantage of the former ones
is much larger critical temperatures while for the latter the lower spatial resolution is
required to observe stationary waves.

\appendix

\section{Collective mode spectrum in a system with different  electron and
hole effective masses in the three-layer dielectric matrix} \label{aa}

Equations (\ref{21})-(\ref{23}) can be easily generalized for $m_e\ne m_h$ and different
$V_{ee}(r)$ and $V_{hh}(r)$.

The functions (\ref{15})-(\ref{17}) written through the Fourier components have the form
\begin{eqnarray}\label{a1}
 \gamma^{(d)}_{\bf k}=V_{ee}(\mathbf{k})\int\frac{d^2 p}{(2\pi)^2}\frac{d^2 p'}{(2\pi)^2}\phi^*_{\mathbf{p}+\tilde{m}_h\mathbf{k}}
\phi_\mathbf{p}\phi^*_{\mathbf{p}'-\tilde{m}_h\mathbf{k}}\phi_{\mathbf{p}'}+V_{hh}(\mathbf{k})\int\frac{d^2
p}{(2\pi)^2}\frac{d^2 p'}{(2\pi)^2}
\phi^*_{\mathbf{p}-\tilde{m}_e\mathbf{k}}\phi_\mathbf{p}
\phi^*_{\mathbf{p}'+\tilde{m}_e\mathbf{k}}\phi_{\mathbf{p}'}\cr+V_{eh}(\mathbf{k})\int\frac{d^2
p}{(2\pi)^2}\frac{d^2 p'}{(2\pi)^2}
\Big[\phi^*_{\mathbf{p}+\tilde{m}_h\mathbf{k}}\phi_\mathbf{p}
\phi^*_{\mathbf{p}'+\tilde{m}_e\mathbf{k}}\phi_{\mathbf{p}'}+
\phi^*_{\mathbf{p}-\tilde{m}_e\mathbf{k}}\phi_\mathbf{p}
\phi^*_{\mathbf{p}'-\tilde{m}_h\mathbf{k}}\phi_{\mathbf{p}'}\Big],
\end{eqnarray}
\begin{eqnarray}\label{a2}
\gamma^{(ex,1)}_{\bf k}=-\int\frac{d^2 p}{(2\pi)^2}\frac{d^2
 q}{(2\pi)^2}
\Big[V_{ee}(\mathbf{p})\Big(\phi^*_{\mathbf{q}} \phi_\mathbf{q}\phi^*_{\mathbf{q}
-\mathbf{p}+\tilde{m}_e\mathbf{k}}\phi_{\mathbf{q}-\mathbf{p}+\tilde{m}_e\mathbf{k}}+
\phi^*_{\mathbf{q}}\phi_{\mathbf{q}-\tilde{m}_h\mathbf{k}}
\phi^*_{\mathbf{q}-\mathbf{p}-\tilde{m}_h\mathbf{k}} \phi_{\mathbf{q}-\mathbf{p}}-
\phi^*_{\mathbf{q}}\phi_{\mathbf{q}} \phi^*_{\mathbf{q}-\mathbf{p}}
\phi_{\mathbf{q}-\mathbf{p}}\Big)\cr +V_{hh}(\mathbf{p})\Big(
\phi^*_{\mathbf{q}}\phi_{\mathbf{q}+\mathbf{p}}
\phi^*_{\mathbf{q}+\mathbf{p}+\tilde{m}_e\mathbf{k}}\phi_{\mathbf{q}
+\tilde{m}_e\mathbf{k}} +
\phi^*_{\mathbf{q}}\phi_{\mathbf{q}+\mathbf{p}-\tilde{m}_h\mathbf{k}}
\phi^*_{\mathbf{q}+\mathbf{p}-\tilde{m}_h\mathbf{k}}\phi_{\mathbf{q}} -
\phi^*_{\mathbf{q}}\phi_{\mathbf{q}+\mathbf{p}}
\phi^*_{\mathbf{q}+\mathbf{p}}\phi_{\mathbf{q} }\Big)\Big]\cr + \frac{1}{2}\int\frac{d^2
p}{(2\pi)^2}\frac{d^2
 q}{(2\pi)^2} V_{eh}(\mathbf{p})
\Big[\phi^*_{\mathbf{q}+\mathbf{p}} \phi_{\mathbf{q}}
\phi^*_{\mathbf{q}+\tilde{m}_e\mathbf{k}} \phi_{\mathbf{q}+\tilde{m}_e\mathbf{k}}+
\phi^*_{\mathbf{q}} \phi_{\mathbf{q}}\phi^*_{\mathbf{q}
+\mathbf{p}+\tilde{m}_e\mathbf{k}}\phi_{\mathbf{q}+\tilde{m}_e\mathbf{k}} +
\phi^*_{\mathbf{q}} \phi_{\mathbf{q}}\phi^*_{\mathbf{q}
+\tilde{m}_e\mathbf{k}}\phi_{\mathbf{q}+\mathbf{p}+\tilde{m}_e\mathbf{k}}\cr +
\phi^*_{\mathbf{q}} \phi_{\mathbf{q}+\mathbf{p}}\phi^*_{\mathbf{q}
+\tilde{m}_e\mathbf{k}}\phi_{\mathbf{q}+\tilde{m}_e\mathbf{k}} +
\phi^*_{\mathbf{q}+\mathbf{p}}\phi_{\mathbf{q}-\tilde{m}_h\mathbf{k}}
\phi^*_{\mathbf{q}-\tilde{m}_h\mathbf{k}}\phi_{\mathbf{q} }
+\phi^*_{\mathbf{q}}\phi_{\mathbf{q}-\tilde{m}_h\mathbf{k}}
\phi^*_{\mathbf{q}+\mathbf{p}-\tilde{m}_h\mathbf{k}}\phi_{\mathbf{q} } +
\phi^*_{\mathbf{q}}\phi_{\mathbf{q}-\tilde{m}_h\mathbf{k}}
\phi^*_{\mathbf{q}-\tilde{m}_h\mathbf{k}}\phi_{\mathbf{q}+\mathbf{p}
}\cr+\phi^*_{\mathbf{q}}\phi_{\mathbf{q}+\mathbf{p}-\tilde{m}_h\mathbf{k}}
\phi^*_{\mathbf{q}-\tilde{m}_h\mathbf{k}}\phi_{\mathbf{q}} -
2\phi^*_{\mathbf{q}+\mathbf{p}}\phi_{\mathbf{q}} \phi^*_{\mathbf{q}}\phi_{\mathbf{q} }
-2\phi^*_{\mathbf{q}}\phi_{\mathbf{q}+\mathbf{p}}
\phi^*_{\mathbf{q}}\phi_{\mathbf{q}}\Big],
\end{eqnarray}
\begin{eqnarray}\label{a3}
 \gamma^{(ex,2)}_{\bf k}=-\int\frac{d^2 p}{(2\pi)^2}\frac{d^2
 q}{(2\pi)^2}\Big[ V_{ee}(\mathbf{p})
\phi^*_{\mathbf{q}-\tilde{m}_e\mathbf{k}} \phi_{\mathbf{q}-\mathbf{k}}\phi^*_{\mathbf{q}
-\mathbf{p}-\tilde{m}_h\mathbf{k}}\phi_{\mathbf{q}-\mathbf{p}}+V_{hh}(\mathbf{p})
\phi^*_{\mathbf{q}-\tilde{m}_e\mathbf{k}}\phi_{\mathbf{q}+\mathbf{p}-\mathbf{k}}
\phi^*_{\mathbf{q}+\mathbf{p}-\tilde{m}_h\mathbf{k}}\phi_{\mathbf{q}
 }\Big]\cr +
\frac{1}{2}\int\frac{d^2 p}{(2\pi)^2}\frac{d^2
 q}{(2\pi)^2} V_{eh}(\mathbf{p})
\Big[\phi^*_{\mathbf{q}+\mathbf{p}-\tilde{m}_e\mathbf{k}} \phi_{\mathbf{q}-\mathbf{k}}
\phi^*_{\mathbf{q}-\tilde{m}_h\mathbf{k}} \phi_{\mathbf{q}}+
\phi^*_{\mathbf{q}-\tilde{m}_e\mathbf{k}} \phi_{\mathbf{q}-\mathbf{k}}\phi^*_{\mathbf{q}
+\mathbf{p}-\tilde{m}_h\mathbf{k}}\phi_{\mathbf{q}}\cr +
\phi^*_{\mathbf{q}-\tilde{m}_e\mathbf{k}}
 \phi_{\mathbf{q}-\mathbf{k}}\phi^*_{\mathbf{q}
-\tilde{m}_h\mathbf{k}}\phi_{\mathbf{q}+\mathbf{p}}+
\phi^*_{\mathbf{q}-\tilde{m}_e\mathbf{k}}
\phi_{\mathbf{q}+\mathbf{p}-\mathbf{k}}\phi^*_{\mathbf{q}
-\tilde{m}_h\mathbf{k}}\phi_{\mathbf{q}}  \Big],
\end{eqnarray}
where $\phi_{\mathbf{q}}$, $V_{ee}(\mathbf{k})$, $V_{hh}(\mathbf{k})$, and
$V_{eh}(\mathbf{k})$  are the Fourier components of the function $\phi_0(\mathbf{r})$,
and of the potentials $V_{ee}(\mathbf{r})$, $V_{hh}(\mathbf{r})$, and
$V_{eh}(\mathbf{r})$, respectively, and we use the notation
$\tilde{m}_{e(h)}={m}_{e(h)}/({m}_e+m_h)$.

For the three-layer dielectric matrix the Fourier components of the Coulomb potentials
can be approximated as
\begin{eqnarray}\label{a4}
V_{ee}({ \mathbf{k}})&=&\frac{4 \pi e^2}{k}
\frac{\varepsilon_2+\varepsilon_3+(\varepsilon_2-\varepsilon_3)e^{-2 k
d}}{(\varepsilon_2+\varepsilon_3)(\varepsilon_2+
\varepsilon_1)-{(\varepsilon_2-\varepsilon_3) (\varepsilon_2-\varepsilon_1)} e^{-2 k
d}},\cr \label{2-1} V_{hh}({ \mathbf{k}})&=&\frac{4 \pi e^2}{k}
\frac{\varepsilon_2+\varepsilon_1+(\varepsilon_2-\varepsilon_1)e^{-2 k
d}}{(\varepsilon_2+\varepsilon_3)
(\varepsilon_2+\varepsilon_1)-{(\varepsilon_2-\varepsilon_3)(\varepsilon_2-\varepsilon_1)}
e^{-2 k d}},\cr
 V_{eh}({ \mathbf{k}})&=&-\frac{8 \pi e^2}{k} \frac{{\varepsilon_2}e^{- k
d}}{{(\varepsilon_2+\varepsilon_3)(\varepsilon_2+\varepsilon_1)}
-{(\varepsilon_2-\varepsilon_3)(\varepsilon_2-\varepsilon_1)} e^{-2 k d}}.
\end{eqnarray}
Equations (\ref{a4}) describe the structure with the following order of layers:
dielectric 1 - $n$-type conducting layer - dielectric 2 - $p$-type conducting layer -
dielectric 3 ($\varepsilon_1$, $\varepsilon_2$, and $\varepsilon_1$  are the dielectric
constants). It is assumed that the thickness of the dielectric layers 1 and 3 is much
larger than $d$.

Below we specify the case  $\varepsilon_2=\varepsilon_3$. The effective Bohr radius of
the pair and the effective Rydberg are given by the relations given above in which the
dielectric constant of the matrix is replaced with
$\varepsilon_{eff}=(\varepsilon_1+\varepsilon_2)/{2}$ ($a_0={\hbar^2
\varepsilon_{eff}}/{m e^2}$, $Ry_{eff}=m e^4/2 \hbar^2 \varepsilon^2_{eff}$). The
spectrum (\ref{19}) can be expressed through dimensionless quantities
\begin{equation}\label{a19}
   E_0(k)=2 Ry_{eff}\sqrt{\left(\tilde{\epsilon}_k+
    [\tilde{\gamma}_{k}^{(ex,1)}-\tilde{\gamma}_{k}^{(ex,2)}]\tilde{n}\right)
    \left(\tilde{\epsilon}_k+
    [2\tilde{\gamma}_k^{(d)}+\tilde{\gamma}_{k}^{(ex,1)}+\tilde{\gamma}_{k}^{(ex,2)}]
    \tilde{n}\right)},
\end{equation}
where $\tilde{\epsilon}_k=\tilde{k}^2 (1-x^2)/8$, $\tilde{k}=k a_0$, $\tilde{n}=n a_0^2$,
and the parameter $x$ is proportional to the difference of the effective masses of
electrons and holes
\begin{equation}\label{a7}
    x=\frac{m_e-m_h}{m_e+m_h}.
\end{equation}
Calculation of the integrals in Eqs. (\ref{a1})-(\ref{a3}) in the limit $d\gg a_0$ yields
the following explicit expressions for $\tilde{\gamma}_k$:
\begin{eqnarray}\label{a9}
 \tilde{\gamma}^{(d)}_{k}=\frac{2 \pi }{\tilde{k}}\left[e^{-\frac{{k}^2 r_0^2
 (1-x)^2}{8}}
 +\frac{\varepsilon_2+\varepsilon_1}{2\varepsilon_2}
 \left(1+\frac{\varepsilon_2-\varepsilon_1}{\varepsilon_2+\varepsilon_1}
 e^{-2 k d}\right)e^{-\frac{{k}^2 r_0^2
 (1+x)^2}{8}}-2e^{-k d} e^{-\frac{{k}^2 r_0^2
 (1+x^2)}{8}}\right],
\end{eqnarray}

\begin{eqnarray}\label{a10}
\gamma^{(ex,1)}_{k}=-\frac{2 \pi r_0}{a_0}
\Bigg[\sqrt{\frac{\pi}{2}}\left(\mathrm{fa}\left[\frac{k^2 r_0^2(1+x)^2}{16}\right]+
e^{-\frac{k^2
 r_0^2(1-x)^2}{8}}-1\right)
\cr
+\sqrt{\frac{\pi}{2}}\frac{\varepsilon_2+\varepsilon_1}{2\varepsilon_2}
\left(\mathrm{fa}\left(\frac{k^2
r_0^2(1-x)^2}{16}\right)+ e^{-\frac{k^2
 r_0^2(1+x)^2}{8}}-1\right)
\cr +\frac{\varepsilon_2-\varepsilon_1}{2\varepsilon_2}\left(e^{-\frac{k^2
 r_0^2(1-x)^2}{8}}\int_0^\infty d p I_0\left(\frac{p k r_0(1-x)}{2}\right)
 e^{-\frac{p^2}{2}-\frac{2 p d}{r_0}}
 +\sqrt{\frac{\pi}{2}}
 e^{\frac{2 d^2}{
 r_0^2}}\mathrm{erfc}\left(\frac{\sqrt{2}d}{ r_0}\right)\left(e^{-\frac{k^2
 r_0^2(1+x)^2}{8}}-1\right)\right)\cr
 -2 e^{-\frac{k^2
 r_0^2(1+x)^2}{8}}\mathrm{fd}\left(\frac{k r_0(1+x)}{4}\right)-2e^{-\frac{k^2
 r_0^2(1-x)^2}{8}}\mathrm{fd}\left(\frac{k r_0|1-x|}{4}\right)+2\mathrm{fd}(0)\Bigg],
 \end{eqnarray}

\begin{eqnarray}\label{a11}
 \gamma_k^{(ex,2)}=-\frac{2 \pi  r_0}{a_0}
\Bigg[\sqrt{\frac{\pi}{2}}e^{-\frac{ k^2 r_0^2(1-x)^2}{8}}\mathrm{fa}\left(\frac{k^2
r_0^2(1+x)^2}{16}\right)+\sqrt{\frac{\pi}{2}}\frac{\varepsilon_2+\varepsilon_1}
{2\varepsilon_2}e^{-\frac{ k^2 r_0^2(1+x)^2}{8}}\mathrm{fa}\left(\frac{k^2
r_0^2(1-x)^2}{16}\right)\cr +\frac{\varepsilon_2-\varepsilon_1}{2\varepsilon_2}
e^{-\frac{k^2
 r_0^2(1+x^2)}{4}}
 \int_0^\infty d p I_0\left(\frac{p k r_0(1-x)}{2}\right)e^{-\frac{p^2}{2}-\frac{2 p d}{r_0}}
 \cr -e^{-\frac{k^2
 r_0^2(1+x^2)}{4}}\left(\mathrm{fd}\left(\frac{k r_0}{2}\right)+\mathrm{fd}\left(\frac{k
 r_0|x|}{2}\right)\right)
\Bigg].
\end{eqnarray}

\begin{figure}
\begin{center}
\includegraphics[width=6cm]{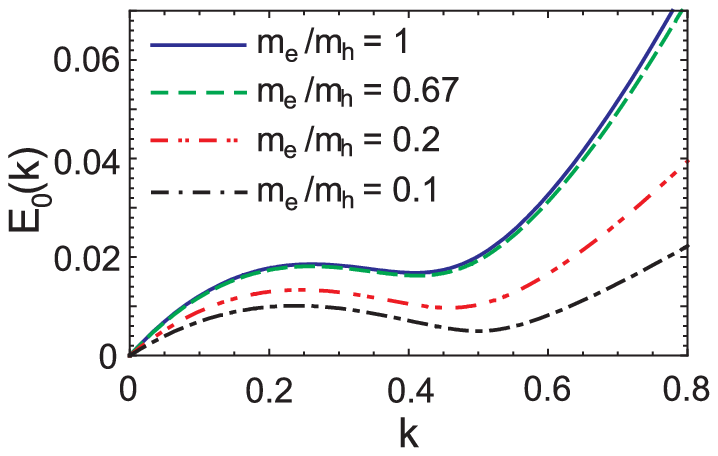}
\end{center}
\caption{Modification of the dispersion curve under variation of $m_e/m_h$ ($n
a_0^2=0.003$, $d=5 a_0$, $\varepsilon_1=\varepsilon_2$). The energy $E_0(k)$ is in units
of $2 Ry_{eff}$, and $k$ is in units of $a_0^{-1}$.}\label{fa1}
\end{figure}

\begin{figure}
\begin{center}
\includegraphics[width=6cm]{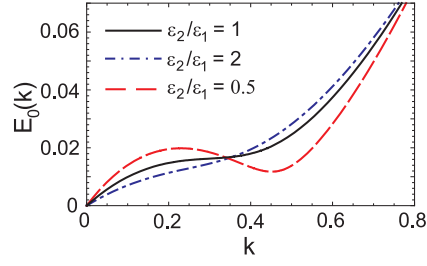}
\end{center}
\caption{Modification of the dispersion curve under variation of
$\varepsilon_1/\varepsilon_2$ ($n a_0^2=0.002$, $d=5 a_0$, $m_e=m_h$). The energy
$E_0(k)$ is in units of $2 Ry_{eff}$, and $k$ is in units of $a_0^{-1}$.}\label{fa2}
\end{figure}

The analysis of Eqs. (\ref{a9})-(\ref{a11}) shows that the roton-type minimum is deeper
in systems where the effective masses of electrons and holes differ from each other (Fig.
\ref{fa1}). It is connected with lowering of the ratio of reduced mass $m$ to the mass of
the pair $m_e+m_h$. The minimum is more shallow at $\varepsilon_2>\varepsilon_1$ and
deeper at $\varepsilon_2<\varepsilon_1$ (Fig. \ref{fa2}). These changes are caused by a
decrease (increase) of the binding potential $V_{eh}$ under variation of
$\varepsilon_2/\varepsilon_1$.

The calculations show that at any $\varepsilon_i$ and $m_e/m_h$ the roton-type minimum
emerges at some $n$ and $d$.  Just the form of the dispersion curve  determines the type
of the stationary wave pattern. Therefore, the patterns shown in Figs. \ref{f2} -
\ref{f7} can be observed at certain  $n$, $d$ and $V$ for any $m_e$ and $m_h$ and for any
$\varepsilon_1$, $\varepsilon_2$, and $\varepsilon_3$. We note that the dispersion curves
for $m_e/m_h = 0.67$  and $m_e/m_h = 1$ almost coincide with each other (see Fig.
\ref{fa1}). It  means that the variation of $m_e/m_h$ in the range $[0.7, 1.5]$ (which
corresponds to TMD systems) does not influence the stationary wave pattern.

\section{Evaluation of the integral in Eq. (\ref{42}).} \label{bb}

The real part of the pole in the integrand in (\ref{42})  is equal to $k_\lambda(\xi)$,
the function defined implicitly by Eq. (\ref{21}). The imaginary part of the pole is
given by the equation
\begin{equation}\label{43}
 k''=\frac{\eta}{\left[\frac{d E_0(k)}{d
k}-\frac{E_0(k)}{k}\right]\Bigg|_{k=k_\lambda(\xi)}}=\frac{\eta}{\left[v_{g0}(\xi)
-v_{p0}(\xi)\right]}.
\end{equation}

\begin{figure}
\begin{center}
\includegraphics[width=6cm]{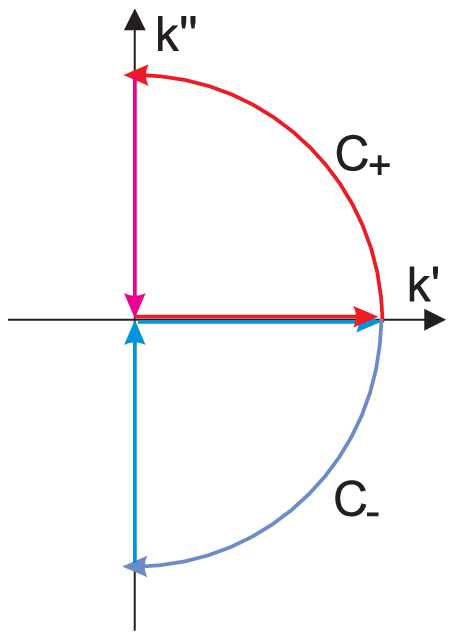}
\end{center}
\caption{Contours $C_+$ (red) and $C_-$ (blue) in the $(k',k'')$ complex plane used for
the calculation of the integral over $k$ in Eq. (\ref{42}). }\label{f8}
\end{figure}

The integral over $k$ in (\ref{42}) can be evaluated using the residue theorem. To apply
this theorem we consider the contours of integration shown in Fig. \ref{f8} and labeled
as $C_+$ and $C_-$. These contours include the integral along the real axis (this is the
integral we are looking for), the integral along the imaginary axis and the integral
along the arc $|k|=\infty$. We choose the contour $C_+$ or $C_-$ depending on the sign of
$\cos(\xi-\chi)$. If $\cos(\xi-\chi)>0$ the integral along the arc is equal to zero for
the contour $C_+$ and we choose the contour $C_+$. In the opposite case
[$\cos(\xi-\chi)<0$] we choose the contour $C_-$. Under such a choice the sum of the
integrals along the real and along the imaginary axes is equal to the residue of the pole
if the latter is located inside the contour and it is equal to zero otherwise. Since the
sign of $k''$ (\ref{43}) coincides with the sign of
$\left[v_{g0}(\xi)-v_{p0}(\xi)\right]$, the relevant contour encircles the pole if
$\cos(\xi-\chi)$ and $\left[v_{g0}(\xi)-v_{p0}(\xi)\right]$ are of the same sign.

The contribution of the residue is proportional to $1/\sqrt{r}$, while the integral along
the imaginary axis is proportional to $1/r^2$ (at $r\gg r_0$). We are interested in the
stationary wave pattern at large distance from the obstacle. Therefore we  neglect the
contribution $\propto1/r^2$ and equate the integral along the real axis to the residue,
\begin{equation}\label{44}
    \delta n(\mathbf{r})=
\frac{n U_0}{\pi}\mathrm{Im}\int_{-\frac{\pi}{2}}^{\frac{\pi}{2}} d\xi \sum_{\lambda}
e^{i k_\lambda(\xi) r
\cos(\chi-\xi)}s_\lambda(\xi,\chi)\left[\frac{k\left[\epsilon_k+(\gamma_k^{(ex,1)}
-\gamma_k^{(ex,2)})n\right]
   }{E_0(k)\left[\frac{d E_0(k)}{d
k}-\frac{E_0(k)}{k}\right]}\right]\Bigg|_{k=k_\lambda(\xi)}.
\end{equation}

The integral over $\xi$ in (\ref{44}) is evaluated in the stationary phase approximation.
The stationary phase condition is
\begin{equation}\label{46}
   \frac{ d f_\lambda(\xi)}{d \xi}=0,
\end{equation}
where the phase $f_\lambda(\xi)$ is given by Eq. (\ref{47}). Equation (\ref{46})
implicitly defines the function $\xi_\lambda(\chi)$. In the general case this function is
the multivalued one. We use the notation $\xi_{\lambda,i}(\chi)$, where $i$ is the branch
number. One can show that the function $\xi_\lambda(\chi)$ is reciprocal to the function
$\chi(\xi)$ defined in Sec. \ref{s3} (see, for instance, \cite{pik1}). By expanding the
phase (\ref{47}) into a series near the stationary points $\xi_{\lambda,i}(\chi)$, we
obtain Eq. (\ref{48}).

\end{document}